# Can Superflares Occur on Our Sun?

Kazunari Shibata[1], Hiroaki Isobe[2], Andrew Hillier[1], Arnab Rai Choudhuri[3,4], Hiroyuki Maehara[1,5], Takako T. Ishii[1], Takuya Shibayama[6], Shota Notsu[6], Yuta Notsu[6], Takashi Nagao[6], Satoshi Honda[1,7], and Daisaku Nogami[1]

[1] Kwasan and Hida Observatories, Kyoto University, Yamashina, Kyoto, 607-8471, Japan
2 Unit of Synergetic Studies for Space, Kyoto University, Yamashina, Kyoto, 607-8471, Japan
3 Department of Physics, Indian Institute of Science, Bangalore, 560012, India
4 National Astronomical Observatory, Mitaka, Tokyo, 181-8588, Japan
5 Kiso Observatory, Institute of Astronomy, School of Science, The University of Tokyo 10762-30, Mitake, Kiso-machi, Kiso-gun, Nagano 397-0101, Japan
6 Department of Astronomy, Faculty of Science, Kyoto University, Kitashirakawa-Oiwake-cho, Sakyo-ku, Kyoto, 606-8502, Japan
7 Nishi-Harima Astronomical Observatory, Center for Astronomy, University of Hyogo, Sayo, Hyogo 679-5313, Japan

Abstract

Recent observations of solar type stars with the Kepler satellite by Maehara et al. have revealed the existence of superflares (with energy of $10^{33} \sim 10^{35}$ erg) on Sun-like stars, which are similar to our Sun in their surface temperature (5600 K ~ 6000 K) and slow rotation (rotational period > 10 days). From the statistical analysis of these superflares, it was found that superflares with energy $10^{34}$ erg occur once in 800 years and superflares with $10^{35}$ erg occur once in 5000 years on Sun-like stars. In this paper, we examine whether superflares with energy of $10^{33} \sim 10^{35}$ erg could occur on the present Sun through the use of simple order-of-magnitude estimates based on current ideas relating to the mechanisms of the solar dynamo. If the magnetic flux is generated by the differential rotation at the base of convection zone as assumed in typical dynamo models, it is possible that the present Sun would generate a large sunspot with total magnetic flux ~2 x $10^{23}$ Mx within one solar cycle period, and lead to superflares with energy of $10^{34}$ erg. On the other hand, it would take ~40 years to store total magnetic flux ~ $10^{24}$ Mx for generating $10^{35}$ erg superflares. Many questions remain relating to how to store $10^{24}$ Mx below the base of convection zone and how to erupt a magnetic flux tube in a short time to create a sunspot with $10^{24}$ Mx, which presents a challenge to dynamo theorists. Hot Jupiters, however, which have been often argued to be a necessary ingredient for generation of superflares, do not play any essential role on generation of magnetic flux in the star itself, if we consider only magnetic interaction between the star and the hot Jupiter. This seems to be consistent with Maehara et al.'s finding of 148 superflare-generating solar type stars which do not have a hot Jupiter companion. Altogether, our simple calculations, combined with Maehara et al.'s analysis of superflares on Sun-like stars, show that there is a possibility that superflares of $10^{34}$ erg would occur once in 800 years on our present Sun, while it is premature to conclude whether it is possible for $10^{35}$ erg superflares to occur on our present Sun on the basis of application of current dynamo theory.

1. Introduction

The solar flare is an explosion generated by magnetic energy release near sunspots in the solar atmosphere (e.g., Shibata and Magara 2011 for a review). The typical amount of energy released in the flare is $10^{29} - 3 \times 10^{32}$ erg (e.g., Priest et al. 1981). Many stars show similar flares (Gershberg 2005), and sometimes the total amount of energy of stellar flares far exceed that of solar flares, say, $10^{33} - 10^{38}$ erg (Shibata and Yokoyama 1999, 2002, Schaefer 1989), especially in young stars and binary stars such as RS CVn. These flares are called superflares (Schaefer et al. 2000).

The first solar flare that human beings observed was the white light flare observed by Carrington (1859) and Hodgson (1859). This flare induced the largest geomagnetic storm (~1760 nT) in the most recent 200 years, and caused damage to the terrestrial telegram system (Loomis 1861, Tsurutani et al. 2003). The total energy of the "Carrington flare" was estimated to be comparable to $10^{32}$ erg on the basis of a sketch of the white light flare (Tsurutani et al. 2003).

In more recent times, the great geomagnetic storm (~ 540 nT) on March 13, 1989 caused a widespread blackout in Quebeck, Canada, and 6 million people had to spend 9 hours without electric power that night. In this case, the flare that led to the geomagnetic storm was X4.6 GOES class in soft X-ray intensity, and the total energy may be also of order of $10^{32}$ erg, considering the energy estimate of other observations of X-class flares (e.g., Benz 2008). Therefore, if superflares with energy more than $10^{33}$ erg would occur on our present Sun, there might be heavy damage to the terrestrial environment and our modern civilization.

Schaefer et al. (2000) reported superflares on ordinary solar type stars (F8 – G8 main sequence stars with slow rotation), but in total only 9 superflares were observed. Hence, it was not possible to discuss statistics in a reliable manner, but they argued that the occurrence frequency of superflares on solar-type stars was of order of once in a few hundred years, where there are no historical records of superflares, or their associated hazards, in the most recent 2000 years. Rubenstein and Schaefer (2000) argued that solar type stars with a hot Jupiter companion are good candidates of superflare stars; i.e., the hot Jupiter may play the role of a companion star in binary stars such as in RS CVn stars that are magnetically very active and produce many superflares. However, there is no hot Jupiter near our Sun, so that Schaefer et al. (2000) predicted that our Sun had never generated superflares and would never produce superflares in future. Schrijver et al. (2012) also argued that flares with energy well above about $10^{33}$ erg are unlikely to occur, considering historical records of sunspot size over the recent 400 years.

Maehara et al. (2012) discovered 365 superflares on solar type stars (G type stars) using Kepler satellite data. Among them, they found 14 superflares on Sun-like stars (slowly rotating G-type main sequence stars, which have rotational periods longer than 10 days and surface temperature of 5600 $\leqq$ $T_{eff}$ < 6000K). From this, they estimated the occurrence frequency of superflares with energy of $10^{34}$ erg is once in 800 years, and that of $10^{35}$ erg superflares is once in 5000 years on Sun-like stars (Fig. 1). If this occurrence frequency of superflares is applicable to our Sun, at some point during the next few thousand years a superflare could lead to heavy damage to the Earth's environment and be hazardous to our modern civilization. This occurrence frequency is comparable to that of the great earthquake which occurred on March 11, 2011 in eastern Japan. Hence, whether superflares would really occur on our Sun is important not only from astrophysical point of view, but also from a social point of view.

In this paper, we will examine whether superflares would occur on our present Sun from a theoretical point of view.

2.  Big Sunspots are Necessary Condition for Superflares

From solar observations, we already know that big flares tend to occur in big sunspot regions. Figure 2 (filled circles) shows the empirical relation between the spot size and the X-ray intensity of solar flares (Sammis et al. 2000, Ishii et al. 2012). If we assume that the X-ray intensity (GOES class) is in proportion to the total released energy by a flare, i.e., if we assume that the energy of the C-class flare is $10^{29}$ erg, M-class $10^{30}$ erg, X-class $10^{31}$ erg, X10-class $10^{32}$ erg, then this result can be interpreted as the evidence that the upper limit of flare X-ray intensity is determined by the scaling law

$$\begin{aligned}
E_{flare} &\approx f E_{mag} \approx f \frac{B^2 L^3}{8\pi} \approx f \frac{B^2}{8\pi} A_{spot}^{3/2} \\
&\approx 7 \times 10^{32} \, [erg] \left(\frac{f}{0.1}\right)\left(\frac{B}{10^3 G}\right)^2 \left(\frac{A_{spot}}{3 \times 10^{19} cm^2}\right)^{3/2} \\
&\approx 7 \times 10^{32} \, [erg] \left(\frac{f}{0.1}\right)\left(\frac{B}{10^3 G}\right)^2 \left(\frac{A_{spot}/(2\pi R_\Theta^{\,2})}{0.001}\right)^{3/2}
\end{aligned} \qquad (1)$$

where f is the fraction of magnetic energy which can be released as flare energy, B is the magnetic field strength, L is the size of the spot, $A_{spot}$ is the area of sunspot, and $R_\Theta$ is the solar radius. The sunspot area for generating X10-class flares (~$10^{32}$ erg) observed in 1989-1997 (Sammis et al. 2000) was $3\times10^{-4}$ of the half area of the solar surface, $\sim10^{19}$ cm$^2$. The total magnetic flux for this case is $10^{22}$ Mx (= G cm$^2$). Using eq. (1), we find that the necessary spot size for generating superflares ($4 \times 10^{33} \sim 10^{35}$ erg) is $0.003 \sim 0.03$ of the half area of the solar surface (Fig. 2), and the necessary total magnetic flux for superflares is $10^{23} - 10^{24}$ Mx.

It is interesting to note that the lifetime of sunspots (T) increases as the area of spots (A) increases (Petrovay and van Driel-Gesztelyi 1977),

$$T \sim A/W, \quad (2)$$

where W ~ 10 MSH/day and MSH = Millionth Solar Hemisphere = $10^{-6} \times 2\pi R_\odot^2 \sim 3.32 \times 10^{16}$ cm$^2$. If this empirical relation holds when extrapolating to very big spots, a lifetime of the superflare generating spot (with magnetic flux $10^{24}$ Mx or area, A $\sim10^{21}$ cm$^2$) becomes $3\times10^3$ days ~ 10 years.

It is interesting to compare our calculation with the frequency distribution of sunspot area. Bogdan et al. (1988) showed that the sunspot area distribution obey log-normal distribution, while Harvey and Zwaan (1993) and more recently Parnell et al. (2009) revealed that the active region area (physically corresponding to total magnetic flux in active region) shows power-law distribution with index of about -2, which is interestingly similar to the flare frequency distribution function. According to these studies, the sunspot with area of $3 \times 10^{19}$ cm$^2$ (corresponding to $3 \times 10^{22}$ Mx) occurs once in a half year. Note that this is an average frequency because no such large spots are observed during the minima of the eleven-year cycle. If the same power-law distribution holds beyond the largest sunspot we observed before, we find the large sunspot with area of $10^{21}$ cm$^2$ (or $10^{24}$ Mx) occurs once in 15 years. This frequency is clearly overestimated; such a large sunspot has never been observed in the last 2 – 3 centuries. On the other hand, if the power-law index is -3 (which may be fitted for area of $3 \times 10^{17} - 10^{18}$ cm$^2$ in Bogdan et al. 1988), the spot with this area occurs once in 1500 years. Thus for power-law index -2 to -3, the frequency of the large spot with magnetic flux of $10^{24}$ Mx (necessary for $10^{35}$ erg superflares) is roughly consistent with the frequency of $10^{35}$ erg superflares.

3. Generation of Magnetic Flux at the Base of the Convection Zone

Is it possible to create $10^{24}$ Mx with the dynamo mechanism of our Sun? Although the current theory of dynamo mechanism has not yet been established, it is generally believed that the magnetic field generation is explained by Faraday's induction equation using the effects of differential rotation and global plasma flow such as global convection or circulation (e.g., Parker 1979, Priest 1982).

Faraday's induction equation is written as (Choudhuri 2003)

$$\frac{\partial \mathbf{B}}{\partial t} = \mathbf{rot}\ (\mathbf{V} \times \mathbf{B})\ , \qquad (3a)$$

$$\frac{\partial B_t}{\partial t} = [\mathbf{rot}\ (\mathbf{V} \times \mathbf{B})]_t \approx B_p R_p \frac{\partial \Omega}{\partial z}\ , \qquad (3b)$$

where $\mathbf{B}$ is the magnetic flux density, $\mathbf{V}$ is the rotational velocity, $\Omega$ is the angular velocity ($V = r\Omega$), $B_t$ is the toroidal component of the magnetic flux density, $B_p$ is the poloidal component of the magnetic flux density, and $R_p$ is the radius of the base of the convection zone. Here the diffusion term is neglected. The condition that the diffusion term can be neglected near the base of the convection zone will be discussed in section 4. The equation (3b) can be integrated in time if the right hand side is constant in time. Then the equation (3b) becomes

$$B_t \approx B_p R_p \frac{\Delta \Omega}{\Delta z} t\ , \qquad (4)$$

where $\Delta \Omega$ is the difference in angular velocity in z-direction between the equator and the pole, $\Delta z$ is the latititudinal thickness of the shear layer of the differential rotation of the Sun. Hence the total magnetic flux generated by the differential rotation in the shear layer (with cross-sectional area of $\Delta r \Delta z$, where $\Delta r$ is the radial thickness of the overshoot‐shear‐layer) may be written as

$$\Phi_t \approx B_t \Delta r \Delta z \approx B_p R_p \frac{\Delta \Omega}{\Delta z} t\ \Delta r \Delta z \approx B_p R_p \Delta \Omega\ t\ \Delta r \approx \Phi_p \frac{\Delta \Omega}{2\pi} t \qquad (5)$$

where $\Phi_p$ is the total poloidal magnetic flux and is given by

$$\Phi_p \approx B_p 2\pi R_p \Delta r\ . \qquad (6)$$

Hence the time scale of generation of the toroidal magnetic flux $\Phi_t$ from the poloidal magnetic flux $\Phi_p$ becomes

$$t \approx \frac{2\pi}{\Delta\Omega} \frac{\Phi_t}{\Phi_p}$$

$$\approx 1.2 \times 10^9 \left(\frac{\Phi_t}{10^{24} Mx}\right) \left(\frac{\Phi_p}{10^{22} Mx}\right)^{-1} \left(\frac{\Delta\Omega}{5.6 \times 10^{-7} Hz}\right)^{-1} \text{sec} \quad (7)$$

$$\approx 40 \left(\frac{\Phi_t}{10^{24} Mx}\right) \left(\frac{\Phi_p}{10^{22} Mx}\right)^{-1} \left(\frac{\Delta\Omega}{5.6 \times 10^{-7} Hz}\right)^{-1} \text{years}$$

Here we used the observed latitudinal differential rotation $\Delta\Omega \sim 0.2 \times \Omega \sim 5.6 \times 10^{-7}$ Hz (e.g., Nandy and Choudhuri 2002, Guerrero and de Gouveia Dal Pino 2007) where $\Omega$ is the present rotation rate $\sim 2.8 \times 10^{-6}$ Hz and assumed

$$\Phi_p \approx B_{polarCH} \pi R_{polarCH}^2 \approx 10 \times 3 \times (0.3 R_\odot)^2 \approx 1 \times 10^{22} \text{ Mx} \quad (8)$$

which is the total poloidal magnetic flux in polar coronal hole, where the average flux density in polar coronal hole is assumed to be $B_{polarCH} = 10$G, and the area of the polar coronal hole is taken to be $\pi R_{polarCH}^2 \approx 3 \times (0.3 R_\odot)^2 \approx$ 1.5 x $10^{21}$ cm². Hence, in order to generate $\Phi_t \sim 10^{24}$ Mx, we need 40 years. This time scale is much shorter than the time interval of the $10^{35}$ erg superflares, i.e. 5000 years. The average generation rate of this magnetic flux during 40 years is 9 x $10^{14}$ Mx/sec. The values of the polar coronal-hole field are used because the polar field of one cycle becomes the source field for the next cycle's toroidal field.

  Though above time scale is longer than the usual cycle length ($\sim$ 11years), it is comparable to the time scale of Maunder Minimum ($\sim$ 70 years). Hence it may be possible to store and increase magnetic flux of $10^{24}$ Mx (for $10^{35}$ erg superflares) below the base of the convection zone for 40 years without having sunspots during that time like in the Maunder Minimum. Within a usual cycle length ($\sim$ 11 years), it would be possible to store magnetic flux 2x$10^{23}$ Mx, so that the superflare with $10^{34}$ erg is more easily produced.

  Observations (e.g., Golub et al. 1974) show that the total magnetic flux emerging for one solar cycle ($\sim$ 11 years) is

$$\Phi \approx B_p S \approx 2 \times 10^{25} \ [Mx] \qquad (9)$$

and the average rate of generation of magnetic flux is $d\Phi/dt \approx 5 \times 10^{16} \ [Mx/\text{sec}]$. This is much larger than the value estimated above. Therefore it is possible to generate the magnetic flux necessary for producing superflares with $10^{35}$ ergs, if generated magnetic flux can be stored for ~ 1 year for above parameters just below the base of the convection zone. However, this observationally estimated generation rate is not necessarily equal to total magnetic flux stored below the base of the convection zone at one time, and hence should be considered to be an upper limit, especially if the flux created in the solar interior is able to make repeated appearances at the surface (Parker 1984).

4. Storage of Magnetic Flux just below the Base of the Convection Zone (Tachocline)

Is it possible to store $10^{24}$ Mx at the base of the convection zone for such a long time (1 ~ 40 years)? How can we reconcile the storage of magnetic flux in the stable region just below the convection zone (overshoot layer) simultaneously with the strong dynamo action due to shear rotation near the Tachocline? These points are the most ambiguous part of present dynamo theory (e.g., Spruit 2012).

The local magnetic flux density at the base of the convection zone is thought to be $3 \times 10^4$ ~ $9 \times 10^4$ G to explain the emerging pattern of sunspot (Choudhuri and Gilman 1987, D'Silva and Choudhuri 1993, Fan et al. 1993). If the possible maximum flux density is assumed to be $10^5$ G at the base of the convection zone (e.g., Ferriz-Mas 1996, Fan 2009), then a flux tube with circular cross-section will have a diameter of order $4 \times 10^9$ cm in order to carry a flux of $10^{24}$ Mx. The question is whether a flux tube of such diameter can be stored at the base of the convection zone for a few years for the toroidal field to be built up by differential rotation.

It has been argued that magnetic flux can be stored within the overshoot layer at the bottom of the convection zone for a long time (van Ballegooijen 1982, Ferriz-Mass 1996). The depth of the overshoot layer has been estimated by various authors (van Ballegooijen 1982, Schmitt et al. 1984, Skaley & Stix 1991) to be a few tenths of pressure scale height (~ $5 \times 10^9$ cm). Skaley & Stix (1991) argued that it can be as thick as 50% of the pressure scale height. Even then it may be difficult to store a flux tube of diameter 4 x 10$^9$ cm entirely within the overshoot layer. However, we get such a large value of the diameter or the vertical extent of the flux tube only if we assume its cross-section to be circular.

We, of course, know that most sunspots are roughly circular. Presumably cross-sections of flux tubes rising through the convection zone become circular due to the twist around them. In fact, it has been argued that flux tubes need to have some twist around them in order to rise as coherent structures (Tsinganos 1980, Cattaneo et al. 1990). Surface observations of sunspots also indicate the presence of helical twist (Pevtsov et al. 1995). However, one of the theoretical models for explaining the helical twists of sunspots (Choudhuri 2003, Choudhuri et al. 2004) suggests that the flux tubes pick up this twist as they rise through the convection zone and the poloidal magnetic field present in the convection zone gets wrapped around them as they rise. If this idea is correct, then there would not be any significant twist around a flux tube at the base of the convection zone and subsequently there would be no reason for the flux tube to have a circular cross-section at the base of the convection zone.

A flattened flux tube with a thickness of 10$^9$ cm in the radial direction can be stored within the overshoot layer without any problem. In order to carry a flux of 10$^{24}$ Mx, such a flux tube will have a latitudinal extension of 10$^{10}$ cm if the magnetic field inside is 10$^5$ G. This latitudinal extension would correspond to a region from the equator to a latitude of about 13 degree at the base of the convection zone. It is certainly not impossible for such a flattened flux tube extending from the equator to 13 degree latitude to be stored within the overshoot layer for several years.

However, there is an energy budget problem (Ferriz-Mas and Steiner 2007). The total kinetic energy of the shear flow due to the differential rotation is estimated to be

$$E_{diff} \approx \frac{\pi}{2} R^2 d \rho_0 v_0^2 \approx 4 \times 10^{37} \ [erg] \quad (10)$$

Here, we assumed

$$R \approx 0.7 \times R_\Theta \approx 5 \times 10^{10} [cm], \ d \approx 10^9 [cm], \ \rho_0 \approx 0.1 [g \, cm^{-3}], \ v_0 \approx 10^4 [cm \, s^{-1}].$$

The total magnetic energy included in a flux tube (with area of $d\Delta z$, where $\Delta z \approx 10^{10} cm$) with total magnetic flux of

$$\Phi \approx d \Delta z B \approx 10^{24} [Mx] \quad (11)$$

is estimated to be

$$E_{mag} \approx 2\pi R d \Delta z \frac{B^2}{8\pi} \approx \frac{R}{4}\Phi B \approx 1.3\times 10^{39}[erg], \qquad (12)$$

which is much larger than the total kinetic energy of differential rotation if B = $10^5$ G is assumed as in the current dynamo theory (Ferriz-Mas and Steiner 2007). Hence the necessary magnetic flux cannot be created by simple stretching of magnetic field lines by shear motion in differential rotation.

Moreno-Insertis, Caligari, and Schuessler (1995), Rempel and Schuessler (2001) and Hotta, Rempel and Yokoyama (2012) have studied a possible intensification mechanism that does not rely on mechanical line stretching through shear motions but utilizes thermal energy : the explosion of rising flux tubes (Ferriz-Mas and Steiner 2007). This mechanism is promising because the thermal energy (~ $10^{13}$ erg cm$^{-3}$) is much larger than kinetic energy (~ $10^7$ erg cm$^{-3}$) at the base of convection zone. However, further studies will be necessary to establish the mechanism to generate $10^5$ G flux tube at the base of the convection zone not only to explain superflares but also to explain normal sunspots.

There is another problem in storage of magnetic flux below the base of the convection zone, i.e., the effect of magnetic diffusivity. In the induction equation (eq. 3), we neglected the effect of diffusion. However, there is an effect of turbulent diffusion in the overshoot layer below the convection zone, because there is turbulence due to overshooting convection. In order to make the flux transport dynamo possible, the advection time for flux transport must be shorter than the diffusion time (Choudhuri, Schuessler and Dikpati 1995). Since the advection time must be shorter than the solar cycle period (~ 10 years) or superflare generating time (~ 40 years), we find

$$R/v \approx t_{ad} < d^2/\eta_{turb} \qquad (13)$$

Hence,

$$\begin{aligned}\eta_{ad} &< d^2/t_{ad} \approx 10^{18}[cm^2]/((3-12)\times 10^8 [sec]) \\ &\approx (0.6-2.4)\times 10^9 [cm^2\ s^{-1}]\end{aligned} \qquad (14)$$

These numbers are comparable to the turbulent diffusivity values assumed in previous flux transport dynamo models (e.g., Hotta and Yokoyama 2010).

Altogether, we conclude that the dynamo mechanism in the present Sun may be able to store $10^{24}$ Mx that can produce superflares of $10^{35}$ erg on the basis of the current idea of typical dynamo model (e.g., Ferriz-Mas 1996, Fan 2009, Choudhuri 2011), though detailed nonlinear processes enabling both generation and storage of $10^{24}$ Mx have not yet been clarified.

5. Case of Rapidly Rotating Stars

It is interesting to note that if the differential rotation rate is in proportion to the rotation rate itself,

$$\Delta\Omega \propto \Omega, \qquad (15)$$

the rate of generation of magnetic flux, i.e., dynamo rate ($f_{dynamo}$) is also in proportion to the rotation rate,

$$f_{dynamo} \approx \frac{d\Phi}{dt}/\Phi \approx \Delta\Omega \propto \Omega \qquad (16)$$

Namely, the dynamo rate becomes larger when the rotation becomes faster. This is consistent with the previous observations that the rapidly rotating stars (such as young stars and RS CVn) are magnetically very active (e.g., Pallavicini et al. 1981, Pevtsov et al. 2003) and show many superflares (e.g., Shibata and Yokoyama 2002). Maehara et al. (2012) also found that the occurrence frequency of superflares in G-type main-sequence stars becomes larger as the rotational period of these stars becomes shorter (see Fig. 3). Figure 3 shows that the occurrence frequency of superflares is roughly inversely proportional to the rotation period (see dashed line). This seems to be consistent with formula (16) if the occurrence frequency of superflares is determined by the generation rate of magnetic flux in stars.

It should be noted that actual observations of differential rotation in late type stars (Barnes et al. 2005, Reiners 2006) revealed behavior different from that assumed here, i.e., the differential rotation decreases with decreasing surface temperature and has a weaker dependence on the rotation rate. However, the evolution of differential rotation in Sun-like stars has not yet been studied well. More detailed study will be necessary for both observations and theories.

6. Is it necessary to have a Hot Jupiter for the production of Superflares ?

On the basis of analogy with RS CVn system, Rubenstein and Schaefer (2000) proposed that "the supeflares occur on otherwise normal F and G main-sequence stars with close Jovian companions, with the superflare itself caused by magnetic reconnection in the field of the primary star mediated by the planet". Ip et al. (2004) studied the star-hot Jupiter interaction, and found that the interaction leads to the energy release via reconnection, which is comparable to that of typical solar flares. Lanza (2008) explained the phase relation between hot spots and the planets within the framework of the same idea. However, it should be noted that in order to generate superflares, strong magnetic field (or large total magnetic flux) must be present in the central star.

Hayashi, Shibata, and Matsumoto (1996) performed magnetohydrodynamic (MHD) simulations of the interaction between a protostar and a disk, and showed that the interaction leads to the twisting of stellar magnetic field and eventual ejection of magnetized plasma, similar to a solar coronal mass ejection (CME), as a result of magnetic reconnection after one rotation of the disk. This model reproduced various observed properties of protostellar flares, which are superflares with energy of $\sim 10^{36}$ erg. Nevertheless, we should remember that the eventual cause of such superflares in the star-disk interaction is the existence of strong magnetic field (or large total magnetic flux) in the central star. If the central star's magnetic flux is small, we cannot have superflares. The differential rotation between the star and the disk can increase the magnetic field strength only to a factor of 2 or 3, because flare/CME occurs soon after one rotation and cannot store more magnetic flux (Hayashi et al. 1996). It is also interesting to note the similarity to solar flares; Moore et al. (2012) shows that a flare occurs when the free magnetic energy (stored near sunspots) is about equal to the energy of the potential magnetic field.

Physically, the role of the hot Jupiter is almost the same as the role of the disk. Therefore we can infer that the hot Jupiter can twist magnetic field between the star and the hot Jupiter only for a short time (~ one orbital period of the hot Jupiter), so that it would be difficult to increase the total magnetic flux of the star itself by magnetic interaction alone. Lanza (2012) reached a similar conclusion considering a more appropriate model for the interaction between the stellar and the planetary magnetic fields.

Cuntz et al. (2000) proposed also tidal interaction between the star and hot Jupiter as the mechanism to enhance magnetic activity. They argued that as a result of tidal interaction, enhanced flows and turbulence in the tidal bulge lead to increased dynamo action. According to their calculation, the tidal force acting on the convection zone of the star by the hot Jupiter is written as

$$F_{tidal} \approx 2 \; \frac{GM_p}{d^2}\frac{R_*}{d} \tag{18}$$

where G is the gravitational constant, $M_p$ is the mass of the planet (hot Jupiter), d is the distance between the star and the planet, and $R_*$ is the radius of the star. Then the ratio of the tidal force to the gravitational force of the star itself is given by

$$\frac{\Delta g}{g} \approx \frac{F_{tidal}}{F_{gravity}} \approx 2 \times \left(\frac{M_p}{M_*}\right)\left(\frac{R_*}{d}\right)^3 \approx 2 \times 10^{-6} \left(\frac{M_p/M_*}{10^{-3}}\right)\left(\frac{R_*/d}{10^{-1}}\right)^3 \quad (19)$$

where $\Delta g$ is the acceleration due to tidal force, and g is the gravitational acceleration of the star itself. This ratio becomes 3 x $10^{-6}$ when $M_p/M_* = 10^{-3}$, and $R_*=R_o=7\times10^{10}$cm, d = 0.04 AU=6x$10^{11}$cm for typical hot Jupiter (Cuntz et al. 2000). It is interesting to note that the ratio of the Coriolis force to gravity force at the base of the solar convection zone is given by

$$\frac{F_{coriolis}}{F_{gravity}} \approx \frac{2V_{conv}\Omega}{GM_\odot/R_\odot^2} \approx 4 \times 10^{-7} \left(\frac{V_{conv}}{10^3 cm/s}\right)\left(\frac{R_\odot \Omega}{2 km/s}\right) \quad (20)$$

Hence the tidal force can be greater than the Coriolis force near the base of the convection zone, so that it may play a role in enhancing the dynamo action there.

In this case, the tidal force may enhance the global convection flow or meridional circulation flow, eventually enhancing the dynamo action, though there is no quantitative calculation. We should remember that the physical mechanism how the tidal force affects the dynamo action has not yet been studied in detail, including Cuntz et al. (2000), and hence the effect of the tidal force on the dynamo action remains vague.

Maehara et al. (2012) did not find hot Jupiters orbiting their 148 superflare stars, but we have to consider the possibility that small planets, invisible to Kepler, may be able to cause a similar tidal interaction. From the above calculation, we can find that the mass of the small planet must be larger than 0.2 $M_J$ (Jupiter mass~ $10^{-3}$ $M_{Sun}$) to cause the effective tidal force to be larger than the normal Coriolis force so that the planet is able to affect the magnetic dynamo activity. If such 0.2 $M_J$ planets are present near superflare stars, they would be detected with the Kepler satellite.

Altogether, we can conclude that the analogy with RS CVn cannot be successfully applied to superflare stars which are slowly rotating but have a hot Jupiter. Namely, the reason of high magnetic activity of RS CVn stars is fast rotation of these stars because of tidal locking. The only possible effect of hot Jupiter on enhancing dynamo action is the tidal interaction, but this argument cannot be applied to superflare stars observed by Maehara et al. (2012) because no exoplanets were observed near superflare stars with the Kepler satellite.

7. Conclusion

We have examined various possibilities relating to whether our Sun can produce superflares, i.e., whether our Sun can generate big sunspot that can lead to occurrence of superflares, using an order-of-magnitude estimate of magnetic flux generation due to typical dynamo mechanism. Although the dynamo mechanism itself has not yet been established, our calculation reveals that it may be possible to generate a big sunspot (with $2 \times 10^{23}$ Mx) that can lead to $10^{34}$ erg superflares within one solar cycle period. On the other hand, we found that it would take 40 years to store magnetic flux ($10^{24}$ Mx) necessary for generating $10^{35}$ superflares. This time scale is much shorter than the time interval (~ 5000 years) for $10^{35}$ superflares, and hence we have enough time to store necessary magnetic flux ($10^{24}$ Mx) below the base of the convection zone within 5000 years. However, we do not know any physical mechanism at present to be able to store such huge magnetic flux by inhibiting emergence of magnetic flux from the base of the convection zone.

It is interesting to note here that this time scale (~40 years) is comparable to Maunder minimum period (~70 years). If we succeed to inhibit the emergence of magnetic flux for more than 40 years, then we may be able to explain both Maunder minimum and $10^{35}$ erg superflares. Large sunspots and superflares are not necessarily occurring at the end of each grand minimum as a result of the field stored during the minimum itself. Specifically, the field could be stored inside the star and then emerge at the surface several years or even decades after the end of the grand minimum. Or it may emerge in several episodes producing many small or medium-sized spots along some time interval. However, it is premature to conclude whether a $10^{35}$ erg superflare could occur on our present Sun on the basis of current dynamo theory. Observations by Maehara et al. (2012) on $10^{35}$ erg superflares on Sun-like stars give a big challenge to current dynamo theory.

We also examined the role of a hot Jupiter production of superflares. Our examination shows that the magnetic interaction alone cannot explain the occurrence of superflares if the total magnetic flux of the central star is small (i.e., comparable to that of the present Sun), whereas the tidal interaction remains to be possible cause of enhanced dynamo activity, though more detailed study would be necessary.

Finally, it is interesting to note that Miyake et al. (2012) recently discovered evidence of strong cosmic rays in $8^{th}$ century (AD 774-775) by analyzing $^{14}C$ data in tree rings in Japan. The cosmic ray-intensity corresponds to a solar flare of energy ~$10^{35}$ erg if the cosmic ray source was a solar flare. Although it may be premature to relate this discovery with superflares on the Sun, it would be interesting to search for evidence of superflares in radio isotopes in tree rings and nitrate ions in antarctic ice cores.

Acknowledgement

Kepler was selected as the tenth Discovery mission. Funding for this mission is provided by the NASA Science Mission Directorate. The data presented in this paper were obtained from the Multimission Archive at STScI. This work was supported by the Grant-in-Aid for the Global COE Program "The Next Generation of Physics, Spun from Universality and Emergence" from the Ministry of Education, Culture, Sports, Science and Technology of Japan. We are grateful to Prof. Kazuhiro Sekiguchi (NAOJ) for useful suggestions.

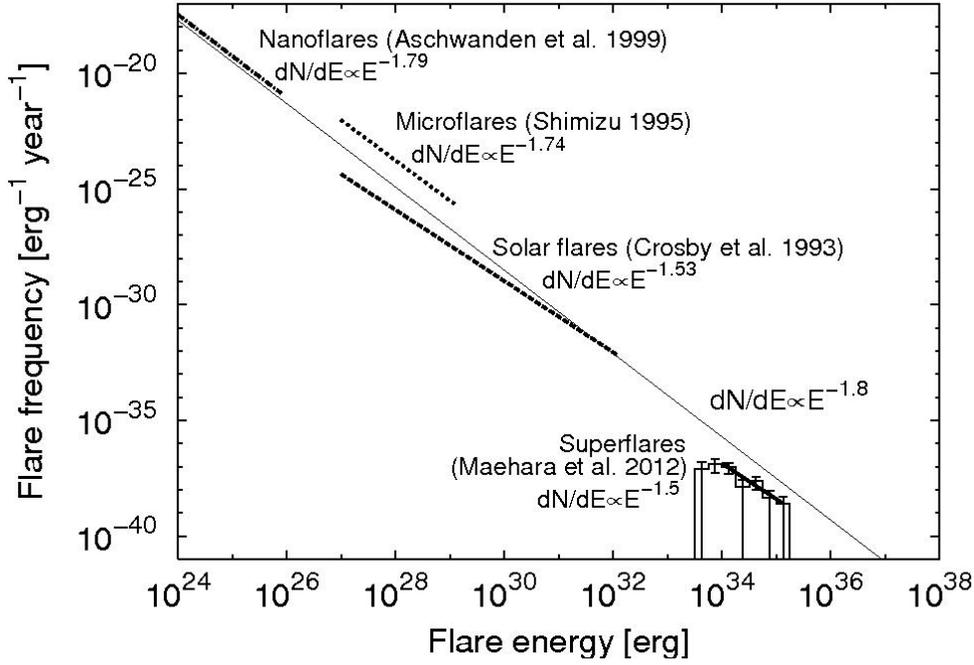

Fig. 1
Comparison between the occurrence frequency of superflares on G-type stars and those of solar flares. The solid-line histogram shows the frequency distribution of superflares on Sun-like stars (slowly rotating G-type main sequence stars with rotational period > 10 days and the effective temperature of 5600-6000K). The error bars in the histogram represent the square root of event number in each bin. This distribution can be fitted by a power-law function with an index -1.5±0.3 (solid line). The dashed line, dotted line and dot-dashed line indicate the power-law distribution of solar flares observed in EUV (Aschwanden et al. 2000), soft X-rays (Shimizu 1995), and hard X-rays (Crosby et al. 1993), respectively. It is interesting that superflares in Sun-like stars, solar flares, microflares, and nanoflares are roughly on the same power-law line with index -1.8 (thin solid line) for wide energy range from $10^{24}$ erg to $10^{35}$ erg.

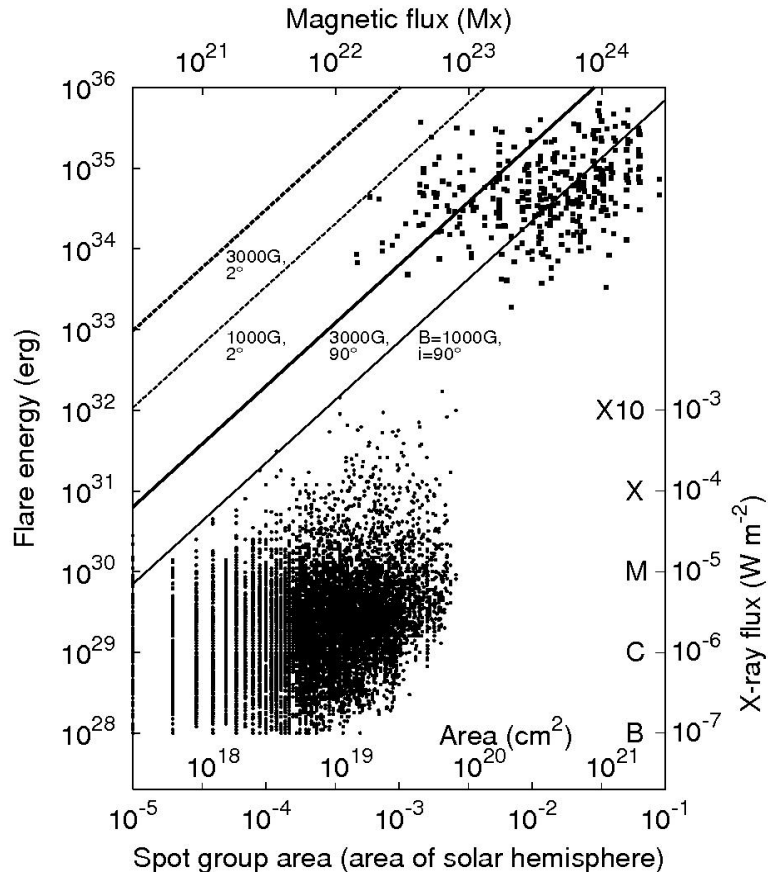

Fig. 2

Flare energy vs sunspot area for superflares on solar type stars (filled squares; Maehara et al. 2012) and solar flares (filled circles; Sammis et al. 2003, Ishii et al. 2012 private communication). The solar flare and sunspot region data are taken from ftp://ftp.ngdc.noaa.gov/STP, and consists of data in 1989-1997 (Sammis et al. 2000) and those in 1996-2006 (Ishii et al. 2012). Thick and thin solid lines corresponds to the analytic relation between the stellar brightness variation amplitude (corresponding to spot area) and flare amplitude (flare energy) obtained from equation (1) (see text) for B=3000G and 1000 G, with i=90 deg and f=0.1, where i is the inclination angle between the rotational axis and the line-of-sight. These lines are considered to give an upper limit for the flare energy (i.e., possible maximum magnetic energy which can be stored near sunspots). However, there are many superflare data points above this line. This is interpreted by Maehara et al. (2012) that these cases may correspond to the stars viewed from above the pole of the rotational axis. That is, in the case of stellar observations, sunspot area is estimated from the apparent stellar brightness variation amplitude, so that if we see stars from the pole, it is not possible to detect star spot. The thick and thin dashed lines correspond to the same relation in case of nearly pole-on ($i$=2.0 deg) for B = 3000 G and 1000 G. Note that the superflare on solar type stars is observed only with visible light and the total energy is estimated from such visible light data. Hence the X-ray intensity in the right hand vertical axis is not based on actual observations. On the other hand, the energy of solar flares are based on the assumption that the energy of X10-class flare is $10^{32}$ erg, X-class $10^{31}$ erg, M-class $10^{30}$ erg, and C-class $10^{29}$ erg, considering previous observational estimate of energies of typical solar flares (e.g., Benz 2008). The values on the horizontal axis at the top show total magnetic flux of spot corresponding to the area on the horizontal axis at the bottom when B = 1000 G.

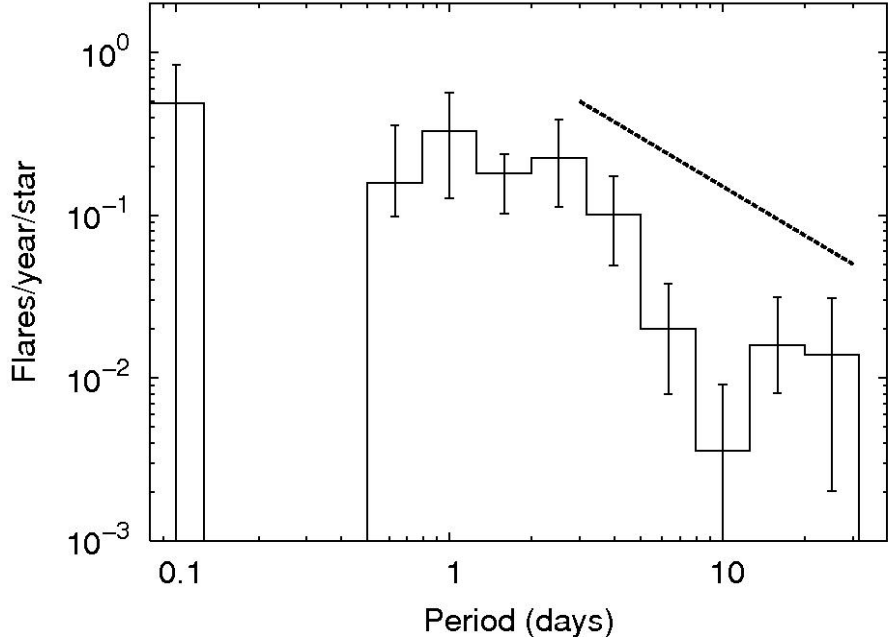

Fig 3  Distribution of occurrence of superflares on G-type main sequence stars in each period bin as a function of variation period (~ rotational period)  (Maehara et al. 2012). The vertical axis indicates the number of flares with the energy ≧ $5\times10^{34}$ erg per 17 star and per year. The error bars represents the $1\sigma$ uncertainty estimated from the uncertainty in the energy estimation and the square root of event number in each period bin. The dashed line shows the line for the occurrence frequency which is in inverse proportion to the rotational period.